\documentstyle[12pt,twoside]{article}
\newcommand{\ba}{\begin{array}}
\newcommand{\ea}{\end{array}}

\newcommand{\A}{{\cal A}}
\newcommand{\B}{{\cal B}}
\newcommand{\C}{{\cal C}}

\newcommand{\Ha}{{\cal H}}

\newcommand{\M}{{\cal M}}

\newcommand{\W}{{\cal W}}

\def\nz{\ifmmode {I\hskip -3pt N} \else {\hbox {$I\hskip -3pt N$}}\fi}
\def\zz{\ifmmode {Z\hskip -4.8pt Z} \else
       {\hbox {$Z\hskip -4.8pt Z$}}\fi}
\def\qz{\ifmmode {Q\hskip -5.0pt\vrule height6.0pt depth 0pt
       \hskip 6pt} \else {\hbox
       {$Q\hskip -5.0pt\vrule height6.0pt depth 0pt\hskip 6pt$}}\fi}
\def\rz{\ifmmode {I\hskip -3pt R} \else {\hbox {$I\hskip -3pt R$}}\fi}
\def\cz{\ifmmode {C\hskip -4.8pt\vrule height5.8pt\hskip 6.3pt} \else
       {\hbox {$C\hskip -4.8pt\vrule height5.8pt\hskip 6.3pt$}}\fi}

\def\au{{\setbox0=\hbox{\lower1.36775ex%
\hbox{''}\kern-.05em}\dp0=.36775ex\hskip0pt\box0}}
\def\ao{{}\kern-.10em\hbox{``}}

\normalsize
\begin{document}
\bibliographystyle{plain}

\begin{titlepage}
\begin{flushright} UWThPh-2002-13\\

\today
\end{flushright}

\vspace*{2.2cm}
\begin{center}
{\Large \bf  On Entanglement of Mesoscopic Systems}\\[30pt]

Heide Narnhofer , Walter Thirring $^\ast $\\ [10pt] {\small\it}
Institut f\"ur Theoretische Physik \\ Universit\"at Wien\\

\vfill \vspace{0.4cm}

\begin{abstract}The entanglement of clouds of N=$10^{11}$ atoms
recently experimentally verified is expressed in terms of the
fluctuation algebra introduced by Verbeure et al. A mean field
hamiltonian describing the coupling to a laser beam leads to
different time evolutions if considered on microscopic or
mesoscopic operators. Only the latter creates non trivial
correlations that finally after a measurement lead to entanglement
between the clouds.

\vspace{0.8cm} PACS numbers: 03.67Hk, 05.30Fk

\\
\hspace{1.9cm}

\end{abstract}
\end{center}

\vfill {\footnotesize}

$^\ast$ {E--mail address: narnh@ap.univie.ac.at}
\end{titlepage}
\vfill \eject

\section{Introduction}

In \cite{JKP} the entanglement of macroscopic objects namely  of
two atomic clouds with $N \sim 10^{11}$ atoms has been observed.
It is not that just a few atoms have been entangled among
themselves but the entanglement took place on the level of a
collective coordinate ${\vec J_N} = \frac {1}{\sqrt N} \sum
_{i=1}^N {\vec j _i}$ if one thinks about the individual atoms as
spins or more generally as angular momenta ${\vec j_i}$. The limit
$N \rightarrow \infty $ of such a quantity was studied in
\cite{GVV1}, \cite{GVV2} and was called fluctuation algebra. It
shows some mathematically tricky aspects and in this paper we
shall investigate their relevance for the phenomenon discovered in
\cite{JKP}.

In a mean field theory one considers the mean magnetizations
${\vec m_N}=\frac {1}{N} \sum _{i=1}^N {\vec j_i}$ which are a
norm bounded set and thus they have weak accumulation points.
Further more  $[m_N,j_i]$ converges in norm to $0$, thus the
accumulation points have to lie in the center of the
representation. This implies that though the existence of $\lim
_{N\rightarrow \infty } {\vec m_N}$ is representation dependent in
an irreducible representation ${\vec m_N}$ must contain
subsequences which converge to a multiple of $1$ (a c-number). In
contradistinction $\|{\vec J_N} \| \sim \sqrt N $ and thus $\vec
J_N$ do not converge in any operator topology. Of course the
unitaries $e^{i{\vec \alpha }({\vec J_N}-<{\vec J_N}
>)}$ are also norm bounded and by the previous argument their weak
accumulation points are also c-numbers. This c-number is the
expectation value of ${\vec m_N}$, thus it is unique in the
representation and corresponds to the measured mean magnetization.
Under appropriate assumptions on the state $\omega $ \cite
{GVV1},\cite{GVV2}, \cite {M1} these limits are $e^{-<{\vec \alpha
}|A|{\vec \alpha }>}$ with some $\omega $-depending $A$ and we
will give an explicit example when the limit is $e^{-|{\vec \alpha
}|^2}$. Thus there is no strong convergence of the unitaries to a
c-number ( which would necessarily have norm $1$). Nevertheless
one can find a state dependent map of the limiting elements onto a
Weyl algebra with a distinguished Gaussian state such that some
properties of expectation values carry over. In particular the
characterization of entanglement as having smaller square
fluctuations than separable  states \cite {Z} also applies to the
limiting algebra.

We consider ${\vec m_N}$ as a macroscopic quantity and call ${\vec
J_N}-<{\vec J_N}>$  mesoscopic in the sense that it is between the
macroscopic and the microscopic level. By the limiting procedure
this mesoscopic quantity is well defined and preserves some of the
quantum structure of the underlying microscopic system.

The tool used in \cite {JKP} for producing entanglement is the
coupling of the ${\vec j_i}$ to a laser beam. This produces a time
evolution $\tau _N^t$  which depends on $N$. For finite $N$ it
certainly exists $\forall t\in R$ but in the limit $N\rightarrow
\infty $ various problems arise. First of all the question arises
whether the limit exists or whether the motion keeps getting
faster. For the microscopic quantities the limit is state
dependent but exists for reasonable states in the same sense as
the mean magnetization does. To carry $\tau _N^t$ over to the Weyl
algebra $\W $ one has to watch out for discontinuities $\lim
_{N\rightarrow \infty} \tau _N^t({\vec J_N})\neq \lim _
{N\rightarrow \infty }\frac {1}{\sqrt N} \sum _{i=1}^N \lim
_{M\rightarrow \infty }\tau _M^t({\vec j_i}).$ The latter exists
under the above conditions, the former also for appropriate states
but is different from the latter. Nevertheless it is the relevant
time evolution ${\tilde \tau _t}$ and in contradistinction to the
other it leads in combination with a measurement of the radiation
field ( and this measurement is necessary!) to entanglement. What
happens is that we have three parties, two clouds of atoms and a
laser beam, that get mixed by ${\tilde \tau _t}$. We start with a
product state, after some time it will not be a product state any
more but nevertheless reduces on the two clouds to a separable
state. If however we measure the radiation field and then reduce
the state it becomes entangled. Clearly the amount of entanglement
depends on the precision of the measurement and only if the
quantum state of the radiation field is completely specified we
can reach optimal entanglement.

\section{The Fluctuation Algebra}

First we repeat the facts that are known about the fluctuation
algebra. It was introduced in \cite{GVV1} and studied in more
detail in \cite{GVV2}. Variations in its definitions adjusted to
varies cluster properties of the underlying state were presented
in \cite{VZ}. Recently the idea was taken up again and the
definition was generalized from strictly local operators to
exponentially localized operators, provided the underlying state
also clusters exponentially \cite{M1},\cite{M2}. The main
definitions and results are the following:

\paragraph{Definition 1:}Let $\A =\overline {\otimes_{l\in Z}
\A_l}$ be an algebra on a one dimensional lattice where the $\A_l
\approx M_d$ are finite dimensional matrix algebras and the
closure is taken in norm. $\A $ contains the local algebra
$\A_{loc} =\bigvee _{\Lambda , |\Lambda |<\infty } \otimes _{l\in
\Lambda } \A_l$. The shift $\alpha^ j :\A_l \rightarrow \A_{l+j}$
is an automorphism of $\A $ as well as of $\A_{loc}$.For shift
invariant states $\omega =\omega \circ \alpha $ and $q\in
\A_{loc}$ we define $$q_{<N>} =\frac{1}{\sqrt{2N+1}} \sum
_{|j|\leq N}(\alpha ^j(q)-\omega (q))$$ $$s(q_1 ,q_2 )=\lim
_{N\rightarrow \infty }\omega (\sum _{|j|\leq N}[q_1, \alpha
^j(q_2)]_-).$$ If $\omega $ is clustering in the sense  that
$$|\omega (q_1 \alpha ^j(q_2))-\omega (q_1) \omega (q_2)|\leq
\|q_1\|\|q_2\|m_j$$ with $\sum_j |j|m_j <\infty $ then we can also
define $$t(q_1 ,q_2 )=\lim _{N\rightarrow \infty }\omega(\sum
_{|j|\leq N} [q_1 -\omega (q_1),\alpha ^j (q_2) -\omega
(q_2)]_+).$$ In the corresponding GNS representation it follows
(\cite{GVV1}, \cite{M1}) that
\paragraph{Proposition:} $\forall q_k \in \A_{loc}, k=1,2,..r$
$$w\lim _{N\rightarrow \infty } \prod e^{iq_{k<N>}} =e^{-i/2\sum
_{k<l}s(q_k,q_l)-\frac {1}{2} t(\sum _k q_k,\sum _l q_l)}$$
\paragraph{Remarks:} The weak convergence refers to the
representation $\Pi _{\omega}$. The weak limit of $e^{iq_{<N>}} $
depends on $\omega $, the strong limit does not exist and
$q_{<N>}$ does not converge even weakly.

The remarkable fact is, that though we only have weak convergence
nevertheless we can assign to the limits an algebraic structure:
\paragraph{Corollary:}To a set $q_k\in \A_{loc},k=1,2...r$ we can
associate by a map $W_{\omega }$ unitaries from a Weyl-algebra $\W
=(e^{iQ_1},e^{iQ_2},..e^{iQ_r})$ with symplectic form
$[Q_k,Q_l]=s(q_k,q_l)$. The state $\omega $ over $\A $ implements
a state $\overline {\omega}$ over $\W $ $$\overline {\omega}(\prod
_{k=1}^r e^{i\beta _k Q_k})=\overline {\omega }(\prod _{k=1}^r
W_{\omega}(q_k))= w\lim _{N\rightarrow \infty } \prod _{k=1}^r
e^{i\beta _k q_{k<N>}}$$ We can illustrate this convergence in the
special
\paragraph{Example:}
We take $\A_j =M_2$ and $\omega (.)=\langle \uparrow
\uparrow...\uparrow|.|\uparrow \uparrow ...\uparrow \rangle.$ Then
$\omega (e^{i\sum \beta _k \sigma _k})=\omega (\cos (|\beta
|)+i\frac {\sum \beta_k \sigma _k}{|\beta |} \sin |\beta |) =\cos
|\beta | +i\beta _z \frac {\sin |\beta |}{|\beta |} $ so that
$$\lim _{\rightarrow \infty }\omega (e^{i(\sum \beta _k \sigma
_k)_{<N>}})=[(\cos \frac{|\beta |}{\sqrt {2N+1}}+i\frac {\beta
_z}{|\beta |}\sin \frac{|\beta |}{\sqrt {2N+1}})e^{-i\frac {\beta
_z}{\sqrt {2N+1}}}]^{2N+1}=$$ $$=\lim _{N\rightarrow \infty
}(1-\frac { |\beta |^2-\beta_ z^2}{2(2N+1)})^{2N+1}
=e^{-1/2(|\beta |^2-\beta _z^2)}$$ which equals the standard
Gaussian state. In this sense $e^{i\beta \sigma_z} \rightarrow 1,
e^{i\beta \sigma _x} \rightarrow e^{i\beta x}, e^{i\beta \sigma
_y} \rightarrow e^{i\beta p}$.

\paragraph{Remark:} Two states lead to the same Weyl algebra resp.
to the same map $W_{\omega }$ if they yield the same $s$ and it
remains to see how far $s$ determines $\omega $. But since in the
definition of the map $W_\omega $ from the quasilocal algebra into
the Weyl algebra the expectation value is included the state on
the Weyl algebra for a given map is unique. Though other states on
the Weyl algebra exist they cannot be constructed from a
quasilocal state by a central limit theorem.

If on the quasilocal algebra $\A $ there exists an automorphisms
$\gamma $ for which $\gamma \A_{loc} \subset \A_{loc} $ then we
can assign to this automorphism an automorphism $\overline {\gamma
}$ on the Weyl-algebra $\W $ by $\overline {\gamma }W_{\omega }(q)
=W_{\omega }( {\gamma } q) $ provided $\omega \circ \gamma =\omega
$ so that the symplectic form remains unchanged. In \cite {M2}
this result is generalized to automorphisms $\gamma $ for which
only $\gamma \A_{exp} \subset \A_{exp} $ i.e. for typical time
evolutions with short range interaction, provided also the state
clusters exponentially. (For the detailed definition see
\cite{M2})

But examining another famous example, the
BCS-model,\cite{T},\cite{GV},we can observe, that the time
evolution on the fluctuation algebra is not inherited from the
time evolution of the quasilocal algebra and ${\overline \gamma }$
may be misleading. As a simpler example we consider the following:

\paragraph {Example:}
We stay in the previous example and consider the time evolution
given by a mean field hamiltonian $$H_N =\frac {1}{N} \sum
_{i,j}^N(\sigma _{x,i} \sigma _{x,j} +\sigma _{y,i} \sigma _{y,j}
+\sigma _{z,i} \sigma _{z,j})$$ For this hamiltonian $$st-\lim
_{N\rightarrow \infty } \frac {d}{dt} \tau _N^t  \sigma _{\alpha
k}=\epsilon ^{\alpha \beta \gamma} m_{\beta }\sigma _{\gamma k},
m_{\beta } =\lim_{N\rightarrow \infty } \sum _{i=1}^N \frac
{\omega (\sigma _{\beta i})}{N} $$ gives on the local level a
rotation of the spins around the mean magnetization and is
therefore state dependent. But for the fluctuation algebra we have
the operator identity $$[H_N, \sum _k \frac {1}{\sqrt N} \sigma
_{x,k} ]= \sum _{i\neq k} \frac {1}{N\sqrt N}(\sigma _{y,i}\sigma
_{z,k} -\sigma _{z,i}\sigma _{y,k})=0.$$ Therefore $\tau
_N^t\sigma _{<N>} =\sigma _{<N>}.$  In the experiment \cite {JKP}
one studies the dynamics $\tau _N$ over $\sigma _{<N>}$ for $N\sim
10^{11}$, therefore in the corresponding limit in the above
example the fluctuation algebra is invariant under the time
evolution.

\section {The Model and its Time Evolution}

In \cite{JKP} two bulks of atoms are considered, independent from
one another in the sense that the state factorizes, but one bulk
is oppositly oriented to the other. The two bulks are influenced
by the same laser beam. By measuring the laser beam after the
interaction the two bulks become entangled. In this paper we want
to study in detail when and how the entanglement emerges. In
\cite{DCJP} the atoms in the bulks together with their interaction
with the laser beam are described in such a way that it is
justified to assign to the laser beam as well as to the bulks
quasilocal algebras on a linear chain where the local algebra of
the laser beam corresponds to the spin algebra describing the
polarization of the laser whereas the local algebra of the bulks
correspond to a finite representation of the angular momentum
describing the different eigenstates of the individual atoms. The
interaction between laser beam and bulk is given by a mean field
hamiltonian. Therefore we are in the framework where we can use
the theory of the fluctuation algebra, but the passage from the
mean field time evolution on the quasilocal algebra to the mean
field time evolution of the fluctuation algebra does not work and
the situation is similar to the BCS-example. The resulting
entanglement resides in long range correlation that are too weak
to be observed on a local level but only emerge in the fluctuation
algebra.

\paragraph{The Model:}
We consider the algebra $\A \otimes \B \otimes \C$ where $$ \A
=\otimes _i\A_i, \A_i =\{\vec {j_i^+}\}$$ $$\B =\otimes _i\B_i,
\B_i=\{\vec {j_i^-}\}$$ $$\C =\otimes _i \C_i, \C_i =\{\vec
{\sigma _i}\}$$ where   ${\vec \sigma _i},{\vec j_k^{\pm }} $ are
sets of independent Pauli matrices (or angular momenta). We assume
that the state factorizes  $\omega =\omega_{\A } \otimes \omega
_{\B } \otimes \omega _{\C }$ with $$\omega _{\B
}(j_{i,y}^+)=\omega _{\B }(j_{i,z}^+)=0$$ $$\omega _{\C
}(j_{i,y}^-)=\omega _{\C } (j_{i,z}^-)=0$$ $$\omega _{\B
}(j_{i,x}^+)=\gamma =-\omega _{\C }(j_{i,x}^-)$$ $$\omega _{\A }
(\sigma _{l,y})=\omega _{\A }(\sigma _{l,z})=0$$ $$\omega _{\A
}(\sigma _{l,x})=s.$$ It is not necessary to specify expectation
values of products, but we assume that the state is exponentially
clustering in space. The influence of the laser beam on the bulks
is given by a hamiltonian $$H_N =\frac{1}{2N} \sum _{l,i,k=1}^N
[a_x\sigma _{l,x} (j_{i,x}^+ +j_{k,x}^-)+a_y\sigma _{l,y}
(j_{i,y}^+ +j_{k,y}^-) +a_z \sigma _{l,z} (j_{i,z}^+
+j_{k,z}^-)]$$ The $a_m\in R$ include the possibility that
$a_x\neq a_y\neq a_z$ so that we cover the time evolution in
\cite{JKP}, but to simplify calculations we will sometimes assume
rotation invariance. For fixed size of the samples, i.e. N finite
and equal for $\A ,\B, \C $ the hamiltonian determines a time
evolution  $$\frac{d\sigma _{l,\alpha }}{dt} =a_{\beta
}\epsilon^{\alpha \beta \gamma }\sigma _{l,\gamma
}\sum_k\frac{j_{k,\beta }^++j_{k,\beta }^-}{N} .$$
$$\frac{dj_{i,\alpha }^+}{dt} =a_{\beta } \epsilon ^{\alpha \beta
\gamma}\sum_l\frac{\sigma _{l,\beta }}{N} j_{i,\gamma }^+$$
$$\frac{dj_{i,\alpha }^-}{dt} =a_{\beta } \epsilon ^{\alpha \beta
\gamma}\sum_l\frac{\sigma _{l,\beta }}{N} j_{i,\gamma}^-.$$ In the
isotropic situation $a_x =a_y =a_z$ the automorphism $\tau_t$ for
fixed $N$ can be written down explicitly. There the total angular
momentum $\sum _k (\sigma _{\alpha ,k}+j_{\alpha ,k}^+ +j_{\alpha
,k}^-)=D_{\alpha }$ is a constant and the motion is a rotation
around it. Exponentiating the operator valued $3 \times 3$ matrix
$(\epsilon D)_{\alpha \beta }=\epsilon _{\alpha \beta \gamma
}D_{\gamma }$ we can write $$\tau _t \sigma _{\alpha ,k}=(e^{\frac
{t}{N}\epsilon D})_{\alpha \beta }\sigma _{\beta },\tau
_tj_{\alpha ,k}^{\pm } =(e^{\frac {t}{N}\epsilon D})_{\alpha \beta
}j_{\beta }^{\pm }$$

The limit $N\rightarrow \infty $ has to be taken with care.

 On the level of the quasilocal algebra we know that $$st-\lim
\sum_l \frac{\sigma _{l,x}}{N}=s$$ $$st-\lim \sum _l\frac{\sigma
_{l,y}}{N} =st-\lim \sum_l\frac{\sigma _{l,z}}{N} =0$$ $$st-\lim
\sum _k\frac{j_{k,x}^+ +j_{k,x}^-}{N} =\gamma -\gamma =0$$ $$
st-\lim\sum _k\frac{j_{k,y}}{N} =st-\lim \sum
_k\frac{j_{k,z}}{N}=0$$ With the iterative solution e.g. $$\tau
_{t,N}^{(n)} \sigma _{l,\alpha }=\sigma _{l,\alpha }+i\int _0^t
dt'[H_N,\tau _{t',N}^{(n-1)}\sigma _{l,\alpha }] $$ we can
conclude $$st-\lim _{N\rightarrow \infty }\tau _{t,N} \sigma
_{l,\alpha }=\sigma _{l,\alpha }, \quad  \alpha =x,y,z $$ $$
st-\lim _{N\rightarrow \infty }j_{k,x}^{\pm } =j_{k,x}^{\pm }$$
$$st-\lim _{N\rightarrow \infty } j_{k,y}^{\pm }=\cos sta_x
j_{k,y}^{\pm } \pm \sin sta_x j_{k,z}^{\pm }$$ and obtain an
automorphism $\tau _t$ that acts on the local algebra and
satisfies $$\omega _{\A } \otimes \omega _{\B } \otimes \omega
_{\C } \circ \tau _t =\omega _{\A } \otimes \omega _{\B } \otimes
\omega _{\C }.$$

 Our assumptions on the state guarantee that we can construct the
fluctuation algebra. With the notation $W_{\omega } (e^{i\alpha
\vec {\sigma }})=e^{i\alpha \vec {S}}$, $W_{\omega }(e^{i\alpha
\vec {j^{\pm }}})=e^{i\alpha \vec {J^{\pm }}}$ the commutation
relations read $$[S_y,S_z]= \sum_j\omega ([\sigma _{i,y}, \sigma _
{j,z} ]) =i\omega (\sigma _{i,z})=is $$ $$[S_y,S_x] =\omega
(\sigma _ {i,z}) =0=[S_z,S_x] $$ and similarly
$$[J_y^+,J_z^+]=i\gamma =[J_z^-,J_y^-]$$ whereas all other
commutators vanish. According to our previous consideration the
time automorphism on the local algebra allows to construct a time
automorphism on the Weyl algebra of fluctuations which acts as a
rotation $$\overline {\tau _t} J_y^{\pm } =\cos sta_x J_y^{\pm }
\pm \sin sta_xJ_z^{\pm }. $$ But this time evolution is not the
one given by $\lim _{\rightarrow \infty } \tau _N^t \sigma
_{<N>}.$ In fact we can rewrite $$\frac {d}{dt}\sum _l \frac
{\sigma _{l,x}-s}{\sqrt {N}} =a_y \sum _l \frac {\sigma
_{l,z}}{\sqrt {N}} \sum _k \frac {j_{k,y}^+ +j_{k,y}^-}{N} -
a_z\sum _l \frac {\sigma _{l,y}}{\sqrt {N}} \sum _k \frac
{j_{k,z}^+ +j_{k,z}^-}{N}$$ Again we can keep in mind that
$$st-\lim \sum \frac {j_{k,x}^+ +j_{k,x}^-}{N} =st-\lim \sum \frac
{j_{k,y}^{\pm }}{N}=st-\lim \sum \frac{j_{k,z}^{\pm }}{N} =0$$
Therefore, using $[H,e^{i\alpha j}]=\int _0^{\alpha }d \alpha
'e^{i\alpha 'j}[H,j]e^{i(\alpha -\alpha ')j}$ we get $$\frac
{d}{dt}\omega (e^{i\alpha \sigma _{x<N>} (t) })=\int _0^{\alpha
}d\alpha ' \omega (e^{i\alpha ' \sigma {x<N>}(t)}\sigma _{z<N>}(t)
\sum \frac {j_{k,y}^+(t) +j_{k,y}^-(t)}{N} e^{i(\alpha -\alpha
')\sigma _{x<N>}(t)})+y\leftrightarrow z$$ with time dependent
operators.  We solve the evolution equation by  using the fact
that the fluctuation algebra is not influenced by local
perturbations so that the strong convergence of the mean values
appearing in the differential equation together with the fact that
$\omega _t(\sigma^2_{z<N>})$ can be controled on the quasilocal
level and is uniformly bounded leads to $$\lim _{N\rightarrow
\infty }\frac {d}{dt}\omega _t(e^{i\alpha \sigma _{x<N>}})=0.$$
For $S_y$ we have to split $$\frac{d}{dt}\sum _l \frac{\sigma
_{ly}}{\sqrt {N}} =-a_x\sum _l\frac {\sigma _{lz}}{\sqrt {N}} \sum
\frac{j_{k,x}^+ +j_{k,x}^-}{N} +a_z\sum _l\frac{\sigma _{l,x}}{N}
\sum _k\frac {j_{kz}^++j_{k,z}^-}{\sqrt {N}}$$ so that we can use
the strong convergence of the mean values we obtain $$\lim
_{N\rightarrow \infty } \omega _t(e^{i\alpha \sigma _{y<N>}})=\lim
_{N\rightarrow \infty } \omega _{\C }(e^{i\alpha \sigma
_{y<N>}})\omega _{\A \otimes \B ,t}(e^{i\alpha
sa_z(j_{z<N>}^++j_{z<N>}^-)})$$ With a similar argument we obtain
$$\frac {d}{dt} \sum _k \frac {j_{k,x} -\gamma }{\sqrt {N}} =a_y
\sum _l \frac {\sigma _{l,y}}{N} \sum _k \frac {j_{k,z}^+}{\sqrt
{N}} -a_z \sum _l \frac {\sigma _{l,z}}{N} \sum _k \frac
{j_{k,y}^+}{\sqrt {N}}=0$$ $$\frac {d}{dt} \sum _k \frac
{j_{k,y}^+}{\sqrt {N}} =-a_x\sum \frac {j_{k,z}^+}{\sqrt {N}} \sum
\frac {\sigma _{l,x}}{N} +a_z\sum  \frac {j_{k,x}^+}{N}\sum \frac
{\sigma _{l,z}}{\sqrt {N} }$$ $$\frac {d}{dt}\sum \frac {j_{k,y}^+
+j_{k,y}^-}{\sqrt {N}} =-a_x \sum \frac {j_{k,z}^+
+j_{k,z}^-}{\sqrt {N}} \sum \frac {\sigma _{l,x}}{N} +a_z \sum
\frac {j_{k,x}^+ +j_{k,x}^- }{N} \sum \frac {\sigma _{l,z}}{\sqrt
{N}}$$ These differential equations define a time evolution
${\tilde \tau _t}$ on the Weyl-algebra $${\tilde \tau _t }S_x
=S_x$$ $${\tilde \tau _t}S_y =S_y +sa_z\int _0^t dt'{\tilde \tau
_t'}(J_z^+ +J_z^-)$$ $${\tilde \tau_t}S_z =S_z -sa_y\int_0^t
dt'{\tilde \tau _t'}(J_y^+ +J_y^-)$$ $${\tilde \tau
_t}J_x^+=J_x^+,{\tilde \tau _t}J_x^- =J_x^-$$ $${\tilde \tau
_t}(J_y^+ +J_y^-)=\cos sta_x (J_y^+ +J_y^-)+\sin st a_x(J_z^+
+J_z^-)$$ $${\tilde \tau _t}(J_z^+ +J_z^-)=-\sin st a_x(J_y^+
+J_y^-) +\cos st a_x (J_z^+ +J_z^-)$$ This can be seen by
evaluating e.g. $$ \lim _{N\rightarrow \infty }\frac {d}{dt}\omega
(e^{iH_N t} e^{i{\tilde \tau _t} \sigma _{x<N>}}e^{-iH_N t})=0$$
Altogether we observe that the time automorphisms ${\tilde \tau }$
and ${\overline \tau }$ coincide on $J^++J^-$ but they do not on
$S$. The automorphism ${\tilde \tau _t}$ is on the level of the
Weyl-algebra implemented by the operator $${\tilde H}=sa_z(J_z^+
+J_z^- )S_z +sa_y(J_y^+ +J_y^-)S_y +\frac {sa_x}{\gamma
}(J_z^++J_z^-)(J_z^+-J_z^-)+\frac {sa_x}{\gamma
}(J_y^++J_y^-)(J_y^+-J_y^-).$$

Generalizing these considerations to other operators of the
Weyl-algebra we first consider as a typical example the operator
$\sum _k \frac {1}{\sqrt N}j_{k,y}^+j_{k+l,z}^+$. If we start with
an even state $\omega $ then for $q_1$ even and $q_2 $ odd
$s(q_1,q_2)$ vanishes. Hence for even $q$ as in our example  the
resulting Weyl-operator commutes with $e^{i\alpha {\vec
S}},e^{i\beta {\vec J^{\pm }}}.$ Neglecting as before terms that
strongly tend to $0$ the time evolution is determined by $$\lim
_{N\rightarrow \infty }\frac {d}{dt}\tau _{t,N} \sum _k\frac
{j_{k,y}^+ j_{k+l,z}^+}{\sqrt N} =\lim _{N\rightarrow \infty
}(sa_x\sum _k \frac {j_{k,z}^+j_{k+l,z}^+}{\sqrt N}-sa_x\sum _k
\frac {j_{k,y}^+j_{k+l,y}^+}{\sqrt N}) =\lim _{N\rightarrow \infty
} \frac {d}{dt} \tau _t\sum _k\frac {j_{k,y}^+ j_{k+l,z}^+}{\sqrt
N}$$ so that this part of the Weyl-algebra inherits the
automorphism of the quasilocal algebra. For another typical
candidate we get $$\lim _{N\rightarrow \infty }\sum \frac
{j_{i,y}^+j_{i+k,y}^+j_{i+l,y}^+}{\sqrt N}=\lim _{N\rightarrow
\infty }(-s\sum \frac {j_{i,z}^+j_{i+k,y}^+j_{i+l,y}^+}{\sqrt
N}+...)+\sum \frac {\sigma _{m,z}}{\sqrt N} \frac
{j_{i,x}^+j_{i+k,y}^+j_{i+l,y}^+}{N}).$$ Here we can use $$w-\lim
_{N\rightarrow \infty} \sum _i \frac
{j_{i,x}^+j_{i+k,y}^+j_{i+l,y}^+}{N}=\lim _{N\rightarrow
\infty}\omega ([\sum _{i,m}\frac
{j_{i,y}^+j_{i+k,y}^+j_{i+l}^+}{\sqrt N},\frac {j_{m,z}^+}{\sqrt
N}])=:\gamma _{kl}$$ with $\gamma _{kl}$ a c- number. If we
therefore calculate the time evolution of the operator $\frac
{j_{i,y}^+j_{i+k,y}^+j_{i+l,y}^+ -\gamma _{kl}j_{i,y}^+}{\sqrt N}
$ that in the limit commutes with ${\vec S}$ and ${\vec J^{\pm }}$
then again the time evolution coincides with the time evolution
inherited from the quasilocal algebra $\overline \tau _t$.
Generalizing our observation we take some $q\in \B_{loc}$ where we
assume, if necessary by replacing $q$ by $q+aj_y^++bj_z^+$ so that
$[Q,{\vec J^+} ]=0$. Then with $q_i =\alpha _i q$ $$\lim \frac
{1}{N}\sum _{i,k} \omega (\left[ [H_N,q_i],{\vec j_k}
\right])=\lim \frac {1}{N} \sum _{i,k} \{\omega (\left[ [{\vec
j_k},H_N],q_i\right ])+\omega (\left [ [q_i, {\vec j_k}
],H_N\right ] ) \}.$$ Now we know on the one hand that $\sum _k
\frac{1}{\sqrt N} [{\vec j_k },H_N]$ remains in this Weyl-algebra
which commutes with the Weyl algebra to which $\sum \frac
{1}{\sqrt N} q_i$ belongs so that the first contribution vanishes.
In the second contribution $[q_i,{\vec j_k } ]$reduces to a local
operator. On local operators we have already proven that the time
evolution corresponds to a local rotation. Together with the
rotation invariance of the state this term vanishes, too.
Altogether the time evolution in the  complement of the Weyl
algebra $\W({\vec S},{\vec J^{\pm } })$ remains in this
complement, and on this complement $\overline \tau ={\tilde \tau
}.$

It remains to interpret the difference of $\overline \tau \neq
{\tilde \tau }$ on $\W({\vec S},{\vec J^{\pm } })$. It results
from the coupling of the parameter $N$ in the mean field
hamiltonian and in the fluctuation algebra. Whether $\overline
{\tau }$ or ${\tilde \tau } $ describes correctly the situation is
determined by the experimental setup. In the experiment in
\cite{JKP} all atoms are influenced by the laser beam and
therefore  we have to choose ${\tilde \tau }$. But this has severe
consequences on the interpretation of the fluctuation algebra.
Whereas for the quasilocal state $\omega _t =\omega \circ \tau _t
= \omega $ now $${\tilde \omega _t} ={\tilde \omega  } \circ
{\tilde \tau _t} \neq {\tilde \omega }.$$ Here we have an example
where the commutation relations determined by $s$ remain
unchanged, whereas different to the considerations in \cite{GVV2}
and \cite{M2} the state evolves in time and will not be reachable
by quasilocal states . If we  want to construct a quasilocal state
that produces ${\tilde \omega _t}$ we fail as can be seen in a
counterexample:

Assume the initial state is a pure product state with $\omega
(\sigma _x ) =\omega (j_x^+ )=-\omega (j_x^- )=1$ and also the
${\vec j}$ are given in a two dimensional representation. Then
$${\overline \omega }(e^{iS_x})={\overline \omega }(e^{iJ_x^+})
={\overline \omega }(e^{-iJ_x^-})=1$$ $${\overline \omega
}(e^{i\alpha S_y})={\overline \omega }(e^{i\alpha S_z})={\overline
\omega }(e^{i\alpha J _y^+})={\overline \omega }(e^{i\alpha J
_z^-})=..=e^{-\alpha ^2}.$$ The expectation values of $S_x$ and
$J_x^{\pm }$ remain unchanged, therefore also the corresponding
state over the quasilocal algebra remains unchanged, which is in
contradiction to ${\tilde \tau _t }(S_y)=S_y+a_t (J_y ^++J_z^-)$
and hence $${\overline \omega _t}(e^{i\alpha S_y})=e^{-\alpha
^2(1+a_t^2)}.$$ We can explain this effect by considering a
sequence of states on the quasilocal algebra, where the
fluctuations of $S_x$ remain of order $1/N$ but the fluctuations
of $S_y$ are larger and are correlated over large distances
according to the long range effect in the mean field hamiltonian.

\section{The Entanglement in the Fluctuation algebra }
Since on the fluctuation algebra our state is Gaussian good
characterizations of entanglement are available\cite{Z}. We are
interested in the entanglement of the tensor product of two
fluctuations algebras resp. two Weyl-algebras. The basic facts are
$$ \delta A^2 +\delta B^2 \geq |<[A,B]>| $$ $$\delta _{\omega
_{\A}\otimes \omega _{\B }}(A\otimes 1+1\otimes B)^2=\delta
_{\omega _{\A }}A^2 +\delta _{\omega _{\B }}B^2$$ If we therefore
consider the variance of our operators ${\vec J_k^{\pm }}$ with
$[J_y^{\pm },J_z^{\pm }]={\pm }i$ in appropriate units, then
$\delta ( J_y^{\pm })^2 +\delta (J_z^{\pm })^2 \geq 1.$ Since by
convex combinations of states the square fluctuations become
greater or equal the convex combinations of the square
fluctuations it follows that in all separable states $$ \delta
(J_y^++J_y^-)^2 +\delta (J_z^++J_z^-)^2\geq \delta (J_y^+)^2+
\delta (J_z^+)^2+ \delta (J_y^-)^2 + \delta (J_z^-)^2 \geq 2.$$
However the general inequality allows that the above fluctuations
of two commuting operators can approach $0$. In order to be sure
that a resulting state is entangled it suffices to calculate that
the fluctuations are sufficiently small. Exactly this
consideration was the basis of the experiment in \cite{JKP} : The
bulk is influenced by a laser beam, and after a measurement on the
laser beam the fluctuations in the bulk are examined and are so
small, that the entanglement of the bulks is proven.

We want to examine in more detail, whether it is the time
evolution or the measurement that is responsible for the
entanglement. To simplify the calculation we assume that we start
with a state that is invariant under the time evolution
${\overline \tau _t}. $ As a consequence $${\tilde \omega }\circ
{\tilde \tau _t} ={\tilde \omega }\circ {\overline \tau _{-t}}
\circ {\tilde \tau _t} ={\tilde \omega } \circ {\hat \tau _t}$$
where $${\hat \tau _t} S_y =S_y+a_t(J_y^+
+J_y^-)+b_t(J_z^++J_z^-)$$ is generated by
$e^{i(J_y^++J_y^-)S_z+i(J_z^++J_z^-)S_y}$ whereas $J_{y,z}^{\pm }$
are ${\hat \tau _t}$ independent. Here $a_t$ and $b_t$ are some
numbers that vary periodically with $t$ and therefore also can
become $0$, but with the appropriate choice of the mean field
hamiltonian also can become arbitrarily large.

 We have  to prove that the time evolution
introduced by the mean field hamiltonian though, as shown before,
mixes the factors, it does not create entanglement in the sense
that taking the partial trace over the laser -algebra $\A$ results
in a state over $\B \otimes \C$, that is not entangled. The state
over $\B \otimes \C $ is determined by the expectation values of
the Weyl operators and we have to remember that the Weyl-algebra
inherits Gaussian states from the local algebra. Therefore we
calculate with omitting or not specifying all unnecessary
parameters $$Tr e^{-i(J_y^+ +J_y^-)S_z -i(J_z^++J_z^-)S_y}\rho
_{\A }\otimes \rho _{\B } \otimes \rho _{\C }e^{+i(J_y^+
+J_y^-)S_z+i(J_z^++J_z^-)S_y}e^{i\alpha J_y^++i\beta J_y^-+i\gamma
J_z^+ +i\delta J_z^-}=$$ $$Tr \rho _{\A }\otimes \rho _{\B }
\otimes \rho _{\C }e^{i\alpha (J_y^++S_z)+i\beta
(J_y^-+S_z)+i\gamma (J_z^+-S_y) +i\delta (J_z^-+S_y)}=$$ $$Tr _{\B
\otimes \C} \rho _{\B }\otimes \rho _{\C }e^{i\alpha J_y^++i\beta
J_y^-+i\gamma J_z^+ +i\delta J_z^-}Tr _{\A }\rho _{\A }e^{i(\alpha
-\beta )S_z  +i(\gamma +\delta )S_y}=$$ $$Tr _{\B \otimes \C} \rho
_{\B }\otimes \rho _{\C }e^{i\alpha J_y^++i\beta J_y^-+i\gamma
J_z^+ +i\delta J_z^-} e^{-c(\alpha -\beta)^2-d(\gamma +\delta
)^2}=$$ $$\int Tr _{\B \otimes \C} \rho _{\B }\otimes \rho _{\C
}e^{i\alpha( J_y^++u)+i\beta ( J_y^--u)-+i\gamma (J_z^++v)
+i\delta (J_z^-+v)} e^{-\frac {1}{4c} u^2-\frac {1}{4d}v^2}
dudv.$$ In this way we have written the expectation value as an
integral over factorizing states and therefore we see that in fact
the time evolution does not produce
 entanglement.

  It remains to examine whether measurements are able
 to destroy separability. First we consider only the maximally
 abelian subalgebra $\M $ generated by
 $S_y,J_y^++J_y^-=:J_y,J_z^++J_z^-=:J_z.$ We start with a product
 state which reduced to $\M $ corresponds to a probability
 distribution that apart from normalization equals
 $e^{-(aS_y^2+J_y^2+J_z^2)}$. Under the time evolution ${\tilde
 \tau _t}$ this becomes $e^{-(aS_y^2+\frac {a_z}{a_x}[(c_t-1)J_y^2+s_tJ_z^2]}$
 (with c and s abbreviating cos and sin). If we now measure $S_y$
 this corresponds to multiplication with a characteristic
 function. For convenience we replace it by a Gaussian $e^{-\frac
 {1}{2}dS_y^2}$ and our probability distribution becomes
 $$e^{-\frac{1}{2}(a+d)+\frac
 {a}{a+d}[..])^2-\frac{ad}{a+d}\frac{a_z^2}{a_x^2}[..]^2-\frac {1}{2}(J_y^2+J_z^2)}.$$
Reduction to $J_y,J_z$ is obtained by integration $\int dS_y$ such
that we finally get $$e^{-\frac {J_y^2}{2}(1+\frac {ad}{a+d}
\frac{a_z^2}{a_x^2}(c_t-1)^2)-\frac {J_z^2}{2}(1+\frac {ad}{a+d}
\frac{a_z^2}{a_x^2}s_t^2)-J_yJ_z\frac {ad}{a+d}\frac{a_z^2}{a_x^2}
(1-c_t)s_t}$$ This probability distribution corresponds to
fluctuations $$\delta J_z^2 =\frac {(a+d)a_x^2
+ada_z^2(c_t-1)^2}{(a+d)a_x^2+2ada_z^2(1-c_t)}$$ $$\delta J_y^2
=\frac {(a+d)a_x^2 +ada_z^2s_t^2}{(a+d)a_x^2+2ada_z^2(1-c_t)}$$
$$\delta J_y^2 +\delta J_z^2=\frac {2(a+d)a_x^2
+2ada_z^2(1-c_t)^2}{(a+d)a_x^2+2ada_z^2(1-c_t)}$$ For $a_x<<a_z$
and $c_t\rightarrow {\pm }1$ the individual fluctuations can be
made arbitrarily small but their sum is always $>1$. This means
that we can go below the limit $2$ and thus generate entanglement,
but the limit $0$ cannot be reached by this kind of measurement.

Nevertheless we know that $J_y$ and $J_z $ commute hence $\delta
J_y^2 +\delta J_z^2 $ arbitrarily small is compatible with the
algebraic structure. Since any function on $J_y$ can only project
on an infinite dimensional subspace we next try whether an optimal
measurement, i.e. a measurement corresponding to a one dimensional
projector can produce arbitrarily small fluctuations for both
$J_y,J_z$. In this case we cannot work only with the maximal
abelian subalgebra but really have to project in the Hilbert space
$\Ha _{\A }.$ We assume in addition that we start with a pure
state over $\A $ given by a Gaussian vector and let it evolve
with$e^{-i{\tilde H}t}$. We start with the Gauss function
$|e^{-\frac {a}{4}J_y^2-\frac {1}{4}(J_y^2 +J_z^2)}\rangle .$ The
$J$ part is invariant under the generator $H_M$ of the microscopic
evolution  so that we can use only the unitaries $e^{-i{\tilde
H}t}e^{iH_M t}.$ Up to an irrelevant phase factor this unitary has
the form $e^{iS_z\frac {a_z}{a_x}[(c_t -1)J_y+s\frac
{a_y}{a_x}[(c_t-1)J_z-s_t J_y]_ +iS_y\frac
{a_y}{a_x}[(c_t-1)J_z-s_t J_y]}.$ It transforms $e^{-\frac
{a}{4}S_y^2}\rightarrow e^{iS_y-\frac {a}{4}(S_y+\frac
{a_z}{a_x}[(c_t -1)J_y+s_t J_z]} .$ Performing a measurement
corresponding to the projector onto a Gauss function $|e^{-\frac
{d}{4}S_y^2} \rangle $ implies that we have to take the square of
the scalar product of the two Gauss function in the $\Ha _{\A }$
space:$$|\int dS_y e^{-\frac {a+d}{4}(S_y+\frac {a}{a+d}\frac
{a_z}{a_x}[(c_t -1)J_y+s_t J_z]+i\frac {1}{a+d}\frac
{a_y}{a_x}[(c_t-1)J_z-s_t J_y])^2} $$ $$e^{-\frac{ad}{a+d}\frac
{a_z^2}{4a_x^2}[(c_t -1)J_y+s_t J_z]^2-\frac
{a_y^2}{4a_x^2(a+d)}[(c_t -1)J_z-s_t J_y]^2+iJm}|^2$$ the
imaginary part drops out and we remain with
$$e^{-\frac{1}{2a_x(a+d)}(ada_z^2[(c_t -1)J_y+s_t
J_z]^2+a_y^2[(c_t-1)J_z-s_t J_y]^2}$$ Together with the rest we
get $e^{-\frac {\alpha }{2}J_z^2-\frac {\gamma }{2}J_y^2 -\beta
J_y J_z } $ where $$\alpha =1+\frac {ada_z^2 s_t^2 +a_y^2
(c_t-1)^2}{a_x^2(a+d)} $$ $$\gamma =1+\frac
{ada_z^2(c_t-1)^2+a_y^2s_t^2}{a_x^2(a+d)}$$ $$\beta =\frac
{s_t(c_t-1)ad(a_z^2+a_y^2)}{a_x^2 (a+d)}$$ Now the square
fluctuations become $$ \delta J_y^2 +\delta J_z^2 =\frac {\alpha +
\gamma }{\alpha \gamma -\beta ^2}$$ $$=\frac
{2a_x^2(a+d)+2(ada_z^2
+a_y^2)(1-c_t)}{a_x^2(a+d)+2(ada_z^2+a_y^2)(1-c_t)+\frac
{4ada_z^2a_y^2c_t^2(1-c_t)^2}{a_x^2(a+d)}}$$
 Since for $a_x
\rightarrow 0$ we can make the last term in the denominator
arbitrarily big in a moment where $c_t\neq 0,1$ the sum of the
square fluctuations can become arbitrarily small.

Collecting these observations we see, that we are in a similar
situation as for the GHZ state \cite{GHZ}$|\uparrow \uparrow
\uparrow +\downarrow \downarrow \downarrow>$: we deal with three
systems, Alice for the laser and Bob and Charles for the two
bulks. The initial pure product state evolves in time to a state
that similar to the GHZ state is no product state any more but
reduced to Bob and Charles is separable ( the tracial state in the
GHZ-example). But  this state can be transformed into a state that
is optimally entangled for Bob and Charles and decoupled from
Alice by a pure local manipulation of Alice (not of Bob and
Charles). For the GHZ state this is$$|\uparrow +\downarrow
><\uparrow +\downarrow |\otimes 1|\uparrow \uparrow \uparrow
+\downarrow \downarrow \downarrow >=|\uparrow +\downarrow
>\otimes |\uparrow \uparrow +\downarrow \downarrow >$$
\section{Microscopic Effects on the Entanglement}
Having created entanglement via a measurement we can still wonder
which are the possible entanglement witnesses and how stable the
entanglement is with respect to the time evolution of the bulk.
the latter question has interesting experimental consequences
\cite {P} .

We have so far localized the entanglement in the Weyl-algebra
$\W(J_y^+,J_z^+,J_y^-,J_z^-).$ From \cite{Z2} we know that the
entanglement can be observed by a violation of a [CHSH] inequality
or by violating \cite{S} the positivity criterium \cite{HHH}.
Enlarging the Weyl-algebra can not increase the entanglement,
because we have already observed that in the Weyl algebra of
fluctuations outside of $\W(J_y^+,J_z^+,J_y^-,J_z^-)$ the mean
field hamiltonian reduces to a strictly local time evolution. Also
on the quasilocal level the entanglement can not be observed,
because here the relevant time evolution $\tau $ does not even
change the state.  Correspondingly  the entanglement sits in long
range correlations, that do not appear on the local level. If we
now assume that on the quasilocal level the time evolution after
the interaction with the laser beam is given by an automorphism
$\gamma _t$ such that $\gamma _t \B =\B $ and $\gamma _t \C =\C $
and $\gamma _t \B_{exp} \subset \B_{exp}$ and satisfying $\omega
\circ \gamma _t =\omega $ and $\gamma _t \circ \alpha ^j =\alpha
^j \circ \gamma _t$ then the symplectic form of the Weyl algebra
is stable under $\gamma _t$. Therefore $\gamma _t$ can be defined
on the Weyl-algebra $${\tilde \gamma _t}W(q) =W(\gamma _t q).$$
${\tilde \gamma _t}$ respects the tensor product structure of the
Weyl-algebra, though it will transfer the entanglement to other
witnesses. Of course we should keep in mind that the state ${\hat
\omega _t} =\omega \circ {\tilde \tau _t} $ does not correspond to
a quasi local state. Therefore the passage from $\gamma $ to
${\tilde \gamma }$ is not justified by the considerations in
chapter $2$. We have to expect that the effect of microscopic time
evolution on the fluctuations might influence  the atypical long
range correlations and could sweep out the entanglement, though it
is not implausible, that this sweeping effect appears only in a
different order of magnitude in time.

\section{Conclusion}
Based on the experiment described in \cite{JKP} and on the
considerations offered there we clarify that the time evolution
has to be expressed on the fluctuation algebra, i.e. on a
mesoscopic level, whereas it does not produce any change of the
state on the microscopic level. As a consequence             the
spatial correlations decay differently than before though not in a
way     that would be observable on a microscopic level. Examining
the time evolved state of the bulks after the interaction with the
laser beam is switched of the state of the two fluctuation bulks
does not factorize  any more but it remains separable. Measuring
the laser beam appropriately   one can produce entanglement of the
two bulks, similar as in the GHZ experiment. Therefore it is not
necessary to expose the two bulks to two different laser beams as
it is done in \cite{JKP}, the second laser beam is only a tool to
observe the entanglement. On the other hand it does not suffice to
expose the bulks to a laser beam, a measurement on the laser beam
is necessary to produce entanglement.
\section{Acknowledgement} We would like to thank A. Verbeure for
drawing our attention to several references and to E.S. Polzik for
informing us about some unpublished results.

\bibliographystyle{plain}

\end{document}